%% file: Crawford.tex
\documentstyle[emulateapj]{article}   

\received{}
\accepted{}
\journalid{}{}
\articleid{}{}

\slugcomment{Accepted by The Astronomical Journal}

\lefthead{Crawford et al.}
\righthead{A Search for Pulsations in FIRST and NVSS Sources}

\newcommand{\approxlt}{\mbox{$\;^{<}\hspace{-0.24cm}_{\sim}\;$}}
\newcommand{\approxgt}{\mbox{$\;^{>}\hspace{-0.24cm}_{\sim}\;$}}

\begin{document}

\title{A Search for Sub-millisecond Pulsations in
Unidentified FIRST and NVSS Radio Sources}

\author{Fronefield Crawford\altaffilmark{1} and Victoria M. Kaspi\altaffilmark{2}} 
\affil{Department of Physics and Center for Space Research,
Massachusetts Institute of Technology, Cambridge, MA 02139, USA}

\and

\author{Jon F. Bell\altaffilmark{3}}
\affil{The University of Manchester, 
Jodrell Bank, Macclesfield,
Cheshire, SK11 9DL, UK}

\altaffiltext{1}{email: crawford@space.mit.edu}
\altaffiltext{2}{email: vicky@space.mit.edu; Alfred P. Sloan Research Fellow} 
\altaffiltext{3}{email: jbell@atnf.csiro.au; current address:
Australia Telescope National Facility, CSIRO, P.O. Box 76, Epping, NSW
2121, Australia}

\begin{abstract}
We have searched 92 unidentified sources from the FIRST and NVSS 1400
MHz radio survey catalogs for radio pulsations at 610 MHz. The
selected radio sources are bright, have no identification with
extragalactic objects, are point-like and are more than 5\% linearly
polarized. Our search was sensitive to sub-millisecond pulsations from
pulsars with dispersion measures (DMs) less than $\sim$ 500 pc
cm$^{-3}$ in the absence of scattering. We have detected no pulsations
from these sources and consider possible effects which might prevent
detection. We conclude that as a population, these sources are
unlikely to be pulsars.
\end{abstract}


\section{Introduction} 

The FIRST survey (Faint Images of the Radio Sky at Twenty Centimeters)
and NVSS survey (NRAO VLA Sky Survey) are recent 1400 MHz VLA radio
surveys of the Northern sky. The FIRST survey is an ongoing survey of
the North and South Galactic caps using the VLA in B-configuration
with a synthesized beam size of $5.4''$ (Becker et al.~1995). In the
published FIRST catalog of radio sources from the first two observing
sessions in 1993 and 1994 (White et al.~1997), 1550 square degrees of
the North Galactic cap were covered spanning $7^{\rm h} <
\alpha$(J2000)$ < 18^{\rm h}$ and +28$^{\circ}$ $<$ $\delta$(J2000)
$<$ +42$^{\circ}$. The positions and flux densities of $\sim 1.4
\times 10^{5}$ discrete radio sources\footnote{The online catalog is
updated regularly as observing proceeds and currently contains more
than $5.4 \times 10^{5}$ sources derived from data taken from 1993 to
1998 (99Jul21 catalog version,
http://sundog.stsci.edu/first/catalogs.html).} are complete down to a
flux density of $\sim$ 1 mJy. The NVSS survey (Condon et al.~1998)
covers $\delta$ $>$ $-40^{\circ}$ (covering 82\% of the celestial
sphere) and catalogs more than 1.8 $\times$ 10$^{6}$ sources complete
down to a flux density of $\sim$ 2.5 mJy. The NVSS survey was
conducted with the VLA in D and DnC configurations with a synthesized
beam size of $45''$. The NVSS survey also preserves polarization
information.

Several large-scale pulsar surveys have previously been conducted at
high Galactic latitudes (see Camilo 1997 and references therein).
However, the rates at which the received analog power was sampled and
digitized in these surveys, typically 3-4 kHz, and the low observing
radio frequencies ($\sim$ 400 MHz), combined with relatively large
radio frequency channel bandwidths of between 125 kHz and 1 MHz,
restricted their sensitivity to sub-millisecond pulsars to very small
dispersion measures (DMs) (DM $\approxlt$ 10 pc
cm$^{-3}$). Large-scale surveys that maintain sensitivity to
sub-millisecond periodicities over a wide range of DMs are difficult:
the fast sampling rate and small radio frequency channel bandwidth
required make large total bandwidths and long integration times
currently impractical. However, a targeted search for sub-millisecond
pulsations is possible using narrow frequency channels and a fast
sampling rate. Such a survey is of course also sensitive to
long-period pulsars which may have been missed in previous surveys due
to radio frequency interference or scintillation.

Consideration of the properties of known recycled pulsars and
representative models of magnetic field decay and equations of state
suggests that a significant population of sub-millisecond pulsars
could be present in the Galaxy (e.g., Possenti et al. 1998). It is
possible, therefore, that some of the sources which remain
unidentified in radio survey catalogs could be bright sub-millisecond
radio pulsars which have previously escaped detection in high-latitude
pulsar surveys. To date, no pulsar has been found with a period
shorter than that of the first millisecond pulsar discovered, PSR
B1937+21, which has a 1.56 ms period (Backer et al. 1982). The
discovery of a sub-millisecond pulsar would place important
constraints on the equation of state of neutron matter at high
densities (e.g., Kulkarni 1992).

\section{Target Choice and Observations}

The FIRST and NVSS surveys contain a number of bright sources which
are unresolved and have no identification in other wavebands.
Although over 99\% of bright sources ($S_{1400} >$ 60 mJy) found in
previous large-scale surveys are believed to be active galactic nuclei
(AGN) (Condon et al. 1998), many sources in the FIRST and NVSS
catalogs remain unidentified. One possibility is that they are
previously unrecognized radio pulsars. Since pulsars often have a high
degree of linear polarization (Lyne \& Manchester 1988), polarized
sources are good targets for pulsar searches. Han \& Tian (1999) have
identified 97 objects in the NVSS catalog which are coincident with
known pulsars. Of the 89 redetected pulsars in Table 1 of their paper
for which the degree of linear polarization could be determined from
the NVSS observations, only 8 had an observed nominal fractional
linear polarization less than 5\%. The intrinsic degree of
polarization of these pulsars may even be higher if bandwidth
depolarization effects are significant, in which case an even higher
fraction of the pulsar sample is more than 5\% linearly
polarized. Figure 2 of Han \& Tian (1999) shows that only $\sim$ 10\%
of identified quasars and $\sim$ 10\% of BL-Lac objects are more than
5\% linearly polarized. Thus, although there is not a clear
polarization cutoff separating the pulsar and extragalactic
populations, a polarization threshold of 5\% excludes most
($\sim$~90\%) of the identified non-pulsar population while retaining
the majority ($\sim$ 90\%) of the identified pulsar population.

We have searched for radio pulsations in bright ($S_{1400} \geq 15$
mJy) point-like unidentified sources from the FIRST and NVSS survey
catalogs which are more than 5\% linearly polarized at 1400
MHz. Sources were selected directly from the catalogs if they met
certain criteria.

Unidentified FIRST sources had their flux densities checked against
their corresponding NVSS flux densities. If a source were extended,
the better resolution of the FIRST survey would be expected to yield a
lower flux density for the source than the NVSS survey. Therefore, in
order to eliminate extended objects, sources were only included if
their FIRST and NVSS flux densities agreed to within a few percent
(indicating an unresolved non-variable source) or if the FIRST flux
density exceeded the NVSS flux density (indicating an unresolved
scintillating source). The large number of extended sources in the
catalogs makes this filter necessary, though it does unfortunately
eliminate scintillating sources which happen to be fainter in the
FIRST survey.

For unresolved NVSS sources outside of the FIRST survey region,
pointed VLA observations were undertaken in October 1995 in
B-configuration in order to obtain the same angular resolution
($5.4''$) as the FIRST survey (R. Becker \& D. Helfand, unpublished
work). Sources in these observations were then subjected to the
selection criteria described above. A total of 92 objects from the
catalogs (39 appearing in both FIRST and NVSS, and 53 appearing only
in NVSS) fit our selection criteria and were sufficiently far north to
be observed in our search. The positions, NVSS total intensity peak
flux densities, and fractional linear polarization from the NVSS peak
flux values are listed for the selected sources in Table \ref{tbl-1}.

Each of the 92 unidentified sources was observed at a center frequency
of 610 MHz in two orthogonal linear polarizations for 420 s with the
Lovell 76-meter telescope at Jodrell Bank, UK. A total bandwidth of 1
MHz was split into 32 frequency channels with detected signals from
each channel added in polarization pairs and recorded on Exabyte tape
as a continuous 1-bit digitized time series sampled at 50 $\mu$s.

The minimum detectable flux density of periodicities in a pulsar
search depends upon the raw sensitivity of the system and a number of
propagation and instrumental effects (e.g., Dewey et
al. 1985). Interstellar dispersion contributes to the broadening of
the intrinsic pulse according to

\begin{equation}
\tau_{\rm DM} = \left( \frac{202}{\nu} \right)^{3} \, {\rm DM} \, \Delta \nu
\end{equation}

\noindent
Here $\tau_{\rm DM}$ is in milliseconds, $\nu$ is the observing
frequency in MHz, DM is the dispersion measure in pc cm$^{-3}$, and
$\Delta \nu$ is the channel bandwidth in MHz. For our observing
system, $\tau_{\rm DM}$ is 1.135 $\mu$s per pc cm$^{-3}$ of DM.  The
fast sampling rate ($t_{\rm samp}$ = 50 $\mu$s) and small channel
bandwidth ($\Delta \nu = 31.25$ kHz) in our survey made it sensitive
to sub-millisecond pulsars for a large range of DMs (DM $\approxlt$
500 pc cm$^{-3}$) in the absence of pulse scattering effects. Our
estimated sensitivity to pulsations for a range of periods and DMs is
shown in Figure \ref{fig-1}. A detailed explanation of the calculation
used to produce Figure \ref{fig-1} can be found in Crawford (2000).

\section{Data Reduction}

In each observation, the frequency channels were dedispersed at 91
trial DMs which ranged from 0 to 1400 pc cm$^{-3}$ and were summed. We
searched this large DM range in the unlikely event that a source could
be a previously missed long-period pulsar with a high DM. For DM
$\approxgt$ 500 pc cm$^{-3}$ the channel dispersion smearing is too
great to maintain sensitivity to sub-millisecond pulsations from all
our sources. However, the Taylor \& Cordes (1993) model of the
Galactic free electron distribution indicates that for all of our
sources, with the exception of two that are within 5$^{\circ}$ of the
Galactic plane, the DMs are expected to be less than 100 pc cm$^{-3}$
regardless of distance. Each resulting dedispersed time series of
$2^{23}$ samples was then coherently Fourier transformed to produce an
amplitude modulation spectrum corresponding to a trial DM.

Radio frequency interference (RFI) produced many false peaks in
certain narrow regions of the modulation spectra at low DMs. We
therefore masked several frequency ranges and their harmonics in which
RFI appeared regularly so that any true pulsar signals were not
swamped by interference. Typically 1-2\% of the modulation spectrum in
each observation was lost in this way.

We then looked for the strongest peaks in the spectra. First, each
modulation spectrum was harmonically summed. In this process, integer
harmonics in the modulation spectrum are summed, enhancing sensitivity
to harmonic signals (e.g., Nice, Fruchter, \& Taylor 1995).  This is
particularly useful for long-period pulsars, which have a large number
of unaliased harmonics. After summing up to 16 harmonic signals, the
highest candidate peaks in the modulation spectrum were recorded along
with the period, DM, and the signal-to-noise ratio (S/N). Redundant
harmonic candidates were then eliminated. Unique candidates were
recorded if they had S/N $>$ 7 and if the candidate period appeared in
at least 10 DM trials. The final candidates for each beam were
inspected by dedispersing the original data at DMs near the candidate
DM and folding the data at periods near the candidate period in order
to look for a broad-band, continuous pulsar-like signal.

This technique was tested by observing several known bright pulsars
(PSR B1937$+$21, PSR B0329$+$54, and PSR J2145$-$0750) throughout the
survey. The results for these pulsars are listed in Table \ref{tbl-2}.
All three pulsars were detected with S/N consistent with our survey
sensitivity, though scintillation affects the detection strengths.

\section{Discussion} 

We did not detect any significant pulsations from the target
sources. Here we consider possible effects which could prevent
detection if they were pulsars.

\begin{figure}
\plotone{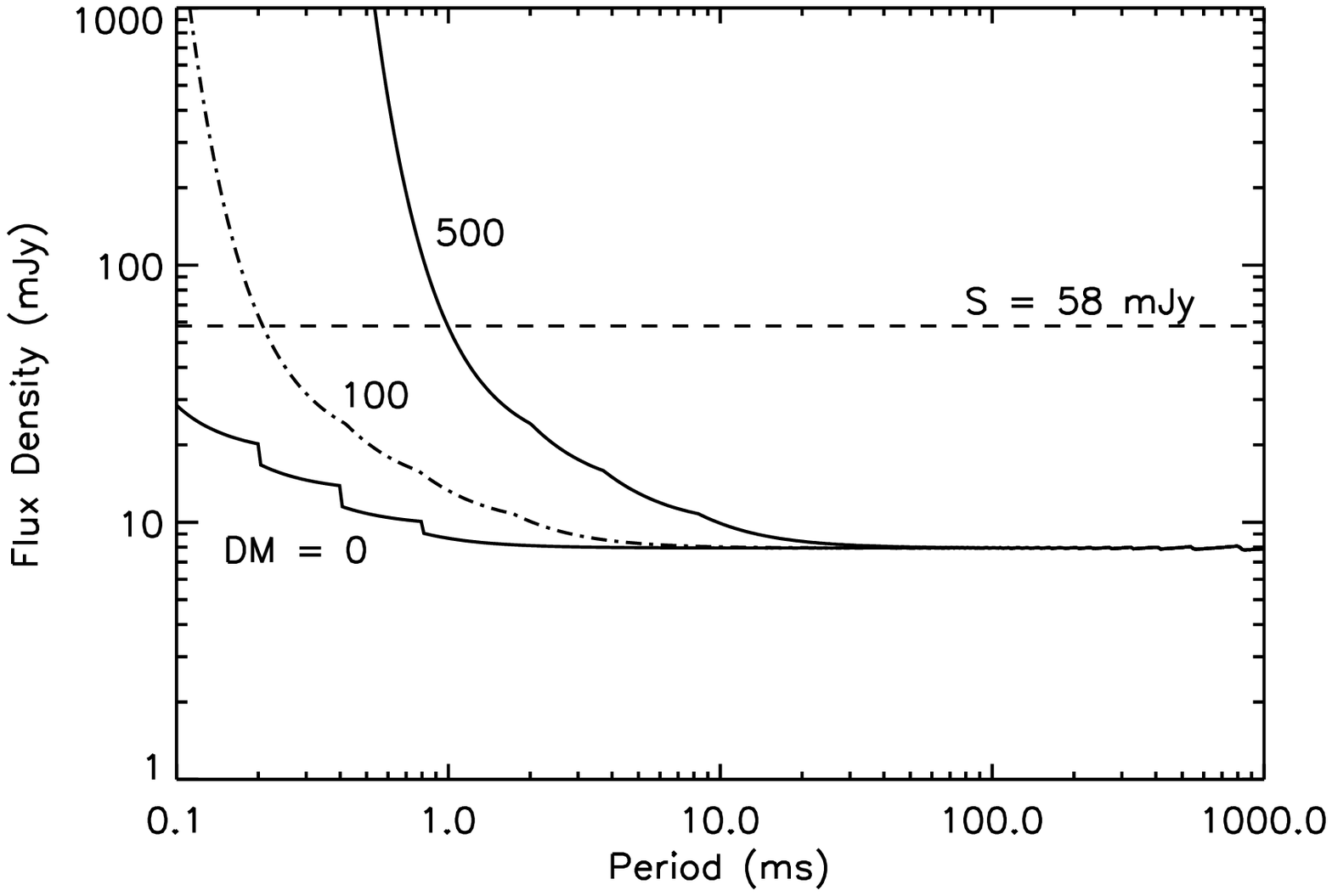} 
\caption{Pulsar sensitivity curves for DMs of 0, 100, and 500 pc
cm$^{-3}$, assuming a 5\% intrinsic pulsed duty cycle. The dashed
horizontal line at 58 mJy is the flux density of our weakest source at
600 MHz, assuming a typical pulsar spectral index of $\alpha = 1.6$.
All of our sources are expected to have DM $<$ 100 pc cm$^{-3}$
(indicated by the dashed-dotted line).}
\label{fig-1} 
\end{figure}

The selected sources were bright, with the weakest source having a
1400 MHz flux density of 15 mJy. Assuming a typical pulsar spectral
index of $\alpha = 1.6$ (Lorimer et al. 1995), where $\alpha$ is
defined according to $S \sim \nu^{-\alpha}$, this source would have
flux density 58 mJy at 600 MHz (the horizontal dashed line in Figure
\ref{fig-1}). For expected DMs, this is about seven (four) times
greater than our sensitivity limit for periods greater than (about
equal to) 1 ms, as indicated in Figure \ref{fig-1}. All of our
sources, therefore, were bright enough (in the absence of
scintillation) to be easily detectable with our observing system if
they were pulsars.

Dispersion smearing is not a factor preventing detection, since all
but two of these sources have high Galactic latitudes ($|b| >
5^{\circ}$) and should have DM $<$ 100 pc cm$^{-3}$ regardless of
distance. This is well within our sensitivity limits to
sub-millisecond pulsations (see Figure \ref{fig-1}). Interstellar
scattering, which can be estimated from the Taylor \& Cordes (1993)
model of the Galactic electron distribution, is expected to be
negligible at 610 MHz. Two of the sources (0458+4953 and 0607+2915)
are within 5$^{\circ}$ of the Galactic plane; for an assumed DM of 100
pc cm$^{-3}$, their predicted pulse scatter-broadening times are
$\sim$ 140 $\mu$s at 610 MHz. This is small enough to maintain
sensitivity to sub-millisecond pulsations, and is of the same order as
the dispersion smearing (113 $\mu$s) at this DM.

An extremely wide beam would prevent modulation of the pulsed signal
and could render a pulsar undetectable. However, this would likely
occur only in an aligned rotator geometry in which both the spin and
magnetic axes are pointing toward us. This is unlikely for our
targets, since, if they were pulsars, the position angle of linear
polarization is expected to follow the projected direction of the
magnetic axis as the star rotates (Lyne \& Manchester 1988). This
geometry would significantly reduce the measured degree of linear
polarization as the pulsar rotates, inconsistent with our choice of
significantly polarized sources.

Scintillation is the modulation of a radio signal passing through a
medium of variable index of refraction, such as an inhomogeneous
interstellar plasma. In diffraction scintillation, the wave scattering
causes interference which can enhance or suppress the amplitude of the
radio signal on the time scale of minutes (e.g., Manchester \& Taylor
1977). These intensity fluctuations vary as a function of radio
frequency at any given time and have a characteristic bandwidth
$\Delta \nu$, where

\begin{equation} 
\Delta \nu \simeq 11 \, \nu^{22/5} \, d^{-11/5} .
\end{equation} 

\noindent
Here $\Delta \nu$ is in MHz, $\nu$ is the observing frequency in GHz,
and $d$ is the distance to the source in kpc (Cordes, Weisberg, \&
Boriakoff 1985). For pulsars with $d <$ 1 kpc, this characteristic
bandwidth exceeds our 610 MHz observing bandwidth of 1 MHz. Indeed, of
28 nearby pulsars observed at 660 MHz by Johnston, Nicastro, \&
Koribalski (1998), 13 had scintillation bandwidths greater than our
bandwidth of 1 MHz and had a characteristic fluctuation time-scale
greater than our integration time of 420 s.  More than half of these
13 pulsars had distances less than 1 kpc, and only one had $d >$ 2
kpc. If the sources we surveyed were placed at a distance of 2 kpc and
a spectral power law index of $\alpha$ = 1.6 is assumed, their 400 MHz
luminosities would all be at the upper end of the observed pulsar
luminosity distribution ($L_{400} > 450$ mJy kpc$^{2}$). This suggests
that if these sources were pulsars, they are likely to be closer than
2 kpc and therefore in the distance range where the scintillation
bandwidth exceeds our observing bandwidth. In this case, scintillation
causes the probability distribution of the observed intensity to be an
exponential function with a maximum in the distribution at zero
intensity (McLaughlin et al. 1999). The number of sources we expect to
see in our sample is the sum of the probabilities that we will see
each individual source (i.e., that the scintillated flux is above the
minimum detectable flux). For the range $P >$ 1 ms, we expect to see
88 of the 92 sources (5\% missed). For the range $P \sim 1$ ms, we
expect to see 85 of the 92 sources (8\% missed).  Thus, only a few of
our sources are likely to have been missed due to
scintillation. Therefore scintillation cannot account for the
non-detection of the bulk of the sources in the survey.

A pulsar in a binary orbit experiences an acceleration which changes
the observed modulation frequency during the course of the
observation. Sensitivity to pulsations is degraded if the frequency
drift exceeds a single Fourier bin $\Delta f = 1 / T_{\rm int}$, where
$T_{\rm int}$ is the integration time. Assuming that the acceleration
of the pulsar is constant during the observation, a critical
acceleration can be defined, above which the change in frequency is
greater than $\Delta f$ and the sensitivity is reduced:

\begin{equation}
a_{\rm crit} = \frac{c}{f T_{\rm int}^2}.
\end{equation} 

\noindent
Since the Fourier drift scales linearly with acceleration, the
reduction in sensitivity also scales linearly with acceleration. For
our integration time of 420 s, the critical acceleration is $a_{\rm
crit} / P$ $\sim 1.7$ (where $a_{\rm crit}$ is in units of m/s$^{2}$
and $P$ is the pulsar period in ms). Our weakest source is several
times brighter than our detection limit (Figure \ref{fig-1}), so a
reduction in sensitivity by a factor of several should still maintain
detectability to pulsations. Thus, a more appropriate critical
acceleration for the weakest source in our list is $a_{\rm crit} / P
\sim 12$ (for $P > 1$ ms) and $a_{\rm crit} / P$ $\sim$ 7 (for $P \sim
1$ ms).

Orbits containing millisecond or sub-millisecond pulsars that have
been spun up from mass transfer from a low-mass donor would be
expected to be circular. Of the 40 known pulsars in circular ($e <$
0.01) binary orbits, the largest projected mean acceleration to our
line of sight (assuming $i=60^{\circ}$) is $a_{\rm mean} / P \sim 3.6$
for PSR J1808$-$3658, an X-ray millisecond pulsar with a 2.5 ms period
in a 2 hr binary orbit around a 0.05 $M_{\odot}$ companion
(Chakrabarty \& Morgan 1998). This acceleration is well below the
critical acceleration $a_{\rm crit} / P$ even for our weakest source.

The mean flux density of the sources on our list, however, is much
higher ($S_{610} \sim 400$ mJy, assuming $\alpha$ = 1.6) than our
weakest source, which raises the critical acceleration for our
typical source. Only very large accelerations ($a_{\rm mean} / P \approxgt 85$
for $P > 1$ ms and $a_{\rm mean} / P \approxgt 40$ for $P \approxlt 1$
ms) would prevent detection of pulsations for our typical source. A
system such as PSR J1808$-$3658 with $S_{610} \sim 400$ mJy would
still be detectable if it had $P \sim 0.3$ ms or if it had an orbital
period $P_{b} \sim 15$ minutes (but not both). Thus it is unlikely
that binary motion in a sub-millisecond or millisecond pulsar system
would be a significant source of non-detections for most of our
sources.

We have also estimated the likelihood of serendipitously detecting a
pulsar not associated with these sources using the observed surface
density of both normal and millisecond pulsars in the Galactic plane
(Lyne et al. 1998). With standard assumptions for a spectral power law
index and luminosity distribution (Lorimer et al. 1993), we find that
it is unlikely ($< 2$\% probability) that we would detect any pulsars
from the chance placement of our 92 beams.

\section{Conclusions}

No pulsations were detected at 610 MHz from the 92 polarized,
point-like sources that we searched from the FIRST and NVSS radio
surveys.  Sensitivity to sub-millisecond pulsations was maintained for
DMs less than about 500 pc cm$^{-3}$ (without scattering effects),
which encompasses the expected DM range for all of these sources.  We
find that several effects which could prevent detection (brightness,
dispersion smearing, scattering, and beaming) are not significant
factors here. Scintillation is expected to account for only a few of
our non-detections and therefore cannot be the cause of the majority
of our non-detections.  For a source with a typical flux density in
our list, Doppler motion in a tight binary system would only prevent
detection if the mean projected line-of-sight acceleration of the
pulsar were at least an order of magnitude higher than those observed
in the known population of circular binary pulsar systems.  We
conclude that as a population, these sources are unlikely to be
pulsars. Given that $\sim$ 10\% of extragalactic sources in the NVSS
survey were identified by Han \& Tian (1999) as being at least 5\%
linearly polarized, it is possible that most of our target sources are
unidentified extragalactic objects. However, the nature of these
sources is still not certain.

\acknowledgments

We thank Bob Becker and David Helfand for providing us with the list
of target sources from the FIRST and NVSS surveys, Andrew Lyne for
assistance with observing, and Nichi D'Amico and Luciano Nicastro for
providing the FVSLAI software package. We thank an anonymous referee
for insightful and helpful comments. VMK is grateful for hospitality
while visiting Jodrell Bank.



\include{Crawford.tab1}

\include{Crawford.tab2}

\newpage

\end{document}

%% file: Crawford.tab1.tex
\begin{deluxetable}{cccccc}
\footnotesize
\tablecaption{Observed Sources \label{tbl-1}}
\tablehead{
\colhead{Source} & 
\colhead{FIRST} & 
\colhead{$\alpha$} &
\colhead{$\delta$} &
\colhead{$S_{1400}$ \tablenotemark{a}} & 
\colhead{\% polarization \tablenotemark{b}} \\
\colhead{} &
\colhead{Source?} & 
\colhead{(J2000)} &
\colhead{(J2000)} &
\colhead{(mJy)}  & 
\colhead{}} 
\startdata
0002$-$1952 & n & 00:02:40.965 & $-$19:52:52.43 & 58  & 9  \nl
0004$-$1148 & n & 00:04:04.905 & $-$11:48:58.52 & 450 & 6  \nl
0011$-$2254 & n & 00:11:09.912 & $-$22:54:58.64 & 37  & 9  \nl
0014$-$2800 & n & 00:14:44.065 & $-$28:00:47.39 & 52  & 14 \nl
0023$-$2155 & n & 00:23:30.215 & $-$21:55:37.73 & 135 & 8  \nl
0024$+$0308 & n & 00:24:49.369 &   +03:08:34.65 & 68  & 9  \nl
0026$-$1112 & n & 00:26:51.454 & $-$11:12:52.57 & 166 & 7  \nl
0027$-$3030 & n & 00:27:02.074 & $-$30:30:32.16 & 24  & 13 \nl
0032$-$2649 & n & 00:32:33.032 & $-$26:49:17.70 & 131 & 6  \nl
0037$-$2323 & n & 00:37:08.808 & $-$23:23:40.65 & 66  & 5  \nl
0040$+$1329 & n & 00:40:21.805 &   +13:29:37.72 & 34  & 9  \nl
0051$+$0229 & n & 00:51:51.304 &   +02:29:44.11 & 15  & 20 \nl
0054$-$1754 & n & 00:54:10.786 & $-$17:54:13.32 & 29  & 11 \nl
0057$+$1341 & n & 00:57:36.448 &   +13:41:45.24 & 64  & 8  \nl
0107$-$1211 & n & 01:07:11.786 & $-$12:11:23.96 & 56  & 6  \nl
0114$-$3219 & n & 01:14:48.887 & $-$32:19:51.76 & 122 & 16 \nl
0138$-$2954 & n & 01:38:40.505 & $-$29:54:46.04 & 45  & 10 \nl
0146$+$0222 & n & 01:46:14.619 &   +02:22:08.16 & 136 & 8  \nl
0147$+$0715 & n & 01:47:27.777 &   +07:15:02.82 & 237 & 6  \nl
0154$-$2422 & n & 01:54:56.898 & $-$24:22:33.61 & 45  & 10 \nl
0214$+$1027 & n & 02:14:59.232 &   +10:27:48.65 & 27  & 12 \nl
0217$-$2354 & n & 02:17:50.767 & $-$23:54:56.42 & 82  & 11 \nl
0223$+$0732 & n & 02:23:33.975 &   +07:32:18.99 & 128 & 12 \nl
0223$+$1159 & n & 02:23:40.829 &   +11:59:10.11 & 34  & 9  \nl
0224$+$1357 & n & 02:24:41.842 &   +13:57:33.00 & 93  & 9  \nl
0238$-$3032 & n & 02:38:55.197 & $-$30:32:02.67 & 155 & 5  \nl
0249$+$1237 & n & 02:49:44.482 &   +12:37:06.27 & 255 & 6  \nl
0251$-$1742 & n & 02:51:06.234 & $-$17:42:39.77 & 65  & 12 \nl
0258$-$3146 & n & 02:58:05.951 & $-$31:46:27.90 & 242 & 8  \nl
0259$+$0747 & n & 02:59:27.067 &   +07:47:39.06 & 807 & 5  \nl
0259$+$4708 & n & 02:59:04.207 &   +47:08:40.31 & 107 & 9  \nl
0317$+$0606 & n & 03:17:26.849 &   +06:06:14.53 & 196 & 8  \nl
0322$-$3458 & n & 03:22:13.098 & $-$34:58:33.34 & 49  & 7  \nl
0326$-$3243 & n & 03:26:15.123 & $-$32:43:24.41 & 93  & 5  \nl
0349$+$0354 & n & 03:49:14.315 &   +03:54:45.34 & 147 & 8  \nl
0403$+$6445 & n & 04:03:42.805 &   +64:45:56.01 & 72  & 5  \nl
0421$+$3511 & n & 04:21:19.710 &   +35:11:15.79 & 68  & 9  \nl
0458$+$4953 & n & 04:58:28.750 &   +49:53:55.67 & 19  & 12 \nl
0505$+$2606 & n & 05:05:54.171 &   +26:06:25.03 & 23  & 9  \nl
0518$+$6439 & n & 05:18:43.662 &   +64:39:57.72 & 28  & 12 \nl
0606$+$4401 & n & 06:06:50.206 &   +44:01:40.73 & 145 & 8  \nl
0607$+$2915 & n & 06:07:18.949 &   +29:15:27.64 & 25  & 14 \nl
0620$+$7334 & n & 06:20:52.108 &   +73:34:41.12 & 84  & 9  \nl
0701$+$2631 & n & 07:01:20.742 &   +26:31:56.95 & 32  & 11 \nl
0719$+$2935 & y & 07:19:22.188 & +29:35:43.30 & 15 & 14  \nl
0733$+$3331 & n & 07:33:13.289 &   +33:31:51.81 & 19  & 13 \nl
0755$+$3013 & y & 07:55:01.887 & +30:13:46.68 & 51 & 11  \nl
0755$+$3341 & y & 07:55:36.599 & +33:41:56.27 & 81 & 7   \nl
0757$+$2721 & y & 07:57:52.648 & +27:21:07.62 & 44 & 7   \nl
0758$+$3929 & n & 07:58:08.846 &   +39:29:28.61 & 530 & 8  \nl
0802$+$3122 & n & 08:02:12.783 &   +31:22:40.56 & 84  & 10 \nl
0805$+$2737 & n & 08:05:19.023 &   +27:37:35.99 & 41  & 9  \nl
0806$+$3310 & n & 08:06:01.704 &   +33:10:10.16 & 44  & 6  \nl
0810$+$3034 & y & 08:10:40.249 & +30:34:32.99 & 152 & 6  \nl
0840$+$2923 & y & 08:40:30.750 & +29:23:32.57 & 17 & 13  \nl
0843$+$3738 & y & 08:43:08.663 & +37:38:16.42 & 108 & 11 \nl
0844$+$3629 & y & 08:44:56.087 & +36:29:27.64 & 49 & 6   \nl
0846$+$3746 & n & 08:46:47.432 &   +37:46:14.97 & 21  & 15 \nl
0903$+$3523 & y & 09:03:05.211 & +35:23:18.91 & 57 & 6   \nl
0911$+$3349 & y & 09:11:47.745 & +33:49:16.60 & 370 & 7  \nl
0923$+$3011 & y & 09:23:30.450 & +30:11:10.92 & 34 & 6   \nl
0928$+$4142 & n & 09:28:22.186 &   +41:42:21.77 & 96  & 12 \nl
0944$+$3803 & n & 09:44:59.202 &   +38:03:17.34 & 42  & 11 \nl
1000$+$3718 & y & 10:00:21.815 & +37:18:44.99 & 35 & 9   \nl
1003$+$3244 & y & 10:03:57.560 & +32:44:02.87 & 419 & 7  \nl
1013$+$3445 & y & 10:13:49.574 & +34:45:50.74 & 350 & 6  \nl
1033$+$2851 & y & 10:33:19.483 & +28:51:22.16 & 29 & 9   \nl
1126$+$3418 & y & 11:26:12.536 & +34:18:20.67 & 39 & 7   \nl
1129$+$3622 & y & 11:29:51.387 & +36:22:15.70 & 119 & 6  \nl
1145$+$3145 & y & 11:45:23.236 & +31:45:17.24 & 77 & 10  \nl
1146$+$2601 & y & 11:46:08.554 & +26:01:05.58 & 114 & 7  \nl
1150$+$3020 & y & 11:50:43.890 & +30:20:17.66 & 31 & 13  \nl
1201$+$2550 & y & 12:01:25.648 & +25:50:04.55 & 20 & 24  \nl
1201$+$3129 & y & 12:01:44.264 & +31:29:03.22 & 87 & 6   \nl
1220$+$3111 & y & 12:20:04.656 & +31:11:45.04 & 28 & 8   \nl
1234$+$2917 & y & 12:34:54.323 & +29:17:43.93 & 434 & 9  \nl
1236$+$3706 & y & 12:36:50.831 & +37:06:02.01 & 62 & 9   \nl
1242$+$2721 & y & 12:42:19.687 & +27:21:57.32 & 69 & 10  \nl
1251$+$3643 & y & 12:51:24.132 & +36:43:57.27 & 29 & 8   \nl
1334$+$3434 & y & 13:34:26.833 & +34:34:25.11 & 48 & 7   \nl
1343$+$2903 & y & 13:43:24.006 & +29:03:57.55 & 22 & 17  \nl
1414$+$4022 & y & 14:14:40.585 & +40:22:25.75 & 43 & 6   \nl
1426$+$4035 & y & 14:26:58.101 & +40:35:38.36 & 28 & 9   \nl
1434$+$3805 & y & 14:34:46.988 & +38:05:14.87 & 149 & 9  \nl
1458$+$3720 & y & 14:58:44.704 & +37:20:22.15 & 210 & 6  \nl
1508$+$2818 & y & 15:08:08.312 & +28:18:13.51 & 76 & 7   \nl
1547$+$3954 & y & 15:47:40.147 & +39:54:38.48 & 128 & 15 \nl
1606$+$2709 & y & 16:06:16.249 & +27:09:28.69 & 29 & 7   \nl
1609$+$2628 & y & 16:09:50.978 & +26:28:38.72 & 17 & 15  \nl
1618$+$2931 & y & 16:18:27.685 & +29:31:17.99 & 29 & 8   \nl
1635$+$3751 & y & 16:35:53.071 & +37:51:54.59 & 47 & 6   \nl
2321$-$1758 & n & 23:21:02.411 & $-$17:58:22.09 & 17  & 13 \nl
\enddata

\tablenotetext{a}{NVSS catalog total intensity 1400 MHz peak flux density.}
\tablenotetext{b}{NVSS catalog percent linear polarization 
from 1400 MHz peak flux densities.} 

\end{deluxetable}

%% file: Crawford.tab2.tex
\begin{deluxetable}{cccccc}
\footnotesize
\tablecaption{Observed Test Pulsars \label{tbl-2}}
\tablehead{
\colhead{Name} & 
\colhead{P} &
\colhead{DM} & 
\colhead{$S_{1400}$ \tablenotemark{a}} & 
\colhead{$S_{600}$ \tablenotemark{b}} & 
\colhead{S/N} \\ 
\colhead{} &
\colhead{(ms)} &
\colhead{(pc cm$^{-3}$)}  & 
\colhead{(mJy)} & 
\colhead{(mJy)} & 
\colhead{} 
} 
\startdata
PSR B0329+54     & 714.52 & 26.8 & 203 & 785 & 725 \nl
PSR B1937+21     &   1.56 & 71.0 &  16 & 100 &  46 \nl
PSR J2145$-$0750 &  16.05 & 9.0  &  10 &  30 &  36 \nl
\enddata

\tablenotetext{a}{Catalog 1400 MHz flux density (Taylor, 
Manchester, \& Lyne 1993).}
\tablenotetext{b}{Catalog 600 MHz flux density (Taylor, 
Manchester, \& Lyne 1993).}

\end{deluxetable}